\documentclass[aps,PRB,twocolumn,superscriptaddress,preprintnumbers]{revtex4-2}
\usepackage[T1]{fontenc}
\usepackage{times}
\usepackage{graphicx,color}
\usepackage{amsfonts,amsmath,amssymb,amsbsy}
\usepackage[colorlinks=true, linkcolor=blue, citecolor=blue, urlcolor=blue]{hyperref}
\usepackage{float}

\begin{document}

\title{Magnetization induced by a nonlinear response to temperature gradient
in $d^{\prime }$, $g^{\prime }$ and $i^{\prime }$ altermagnets}
\author{Motohiko Ezawa}
\affiliation{Department of Applied Physics, The University of Tokyo, 7-3-1 Hongo, Tokyo
113-8656, Japan}

\begin{abstract}
It is a highly nontrivial question whether magnetization is induced by a
nonlinear response to a temperature gradient in systems where the linear
response is forbidden by inversion symmetry. In this work, we address this
issue and provide explicit examples demonstrating that such a response can
indeed arise. The spin-split electron band structures induced by $d$-wave, $%
g $-wave, $i$-wave altermagnets are characterized by $k^{N_{X}}\sin
N_{X}\phi $ around the band bottom, where $N_{X}=2,4$ and $6$, respectively.
In contrast, those of the corresponding $d^{\prime }$-wave, $g^{\prime }$%
-wave, $i^{\prime }$-wave altermagnets are described by $k^{N_{X}}\cos
N_{X}\phi $. We show that a finite magnetization is induced in the $%
d^{\prime }$-wave, $g^{\prime }$-wave, $i^{\prime }$-wave altermagnets under
a second-order nonlinear response to a temperature gradient  $(\partial
_{x}T)^2$  when the temperature gradient is  linear and  along
either the $x$ or $y$ direction, whereas no such response occurs in the $d$%
-wave, $g$-wave and $i$-wave altermagnets. We also study the case where the
direction of the temperature gradient is arbitrary. In this case, a finite
magnetization is induced also in $d$-wave altermagnets but not in $g $-wave
and $i$-wave altermagnets. We discuss the difference in physics between the $%
X$-wave and $X^{\prime }$-wave altermagnets with $X=d,g,i$ based on the
tight-binding model.  Furthermore, we show the emergence of the
magnetization as a response when the temperature profile is not linear, $%
\partial _{x}^{2}T\neq 0$. 
\end{abstract}

\date{\today }
\maketitle

\section{Introduction}

Nonlinear responses have attracted considerable attention. The most studied
one is the nonlinear electric conductivity induced by a nonlinear response
to electric field\cite%
{Gao,HLiu,Michishita,Watanabe,CWang,Oiwa,AGao,NWang,KamalDas,Kaplan,Ohmic,Xiang,EzawaMetricC,YFang,EzawaPNeel}%
. The nonlinear spin conductivity has also been studied\cite%
{Hamamoto,Kameda,Hayami22B,Hayami24B,GI,HJ}. A nonlinear response to a
temperature gradient $\nabla _{x}T$ can generate currents including the
charge current\cite{Yu,Karki,Zeng,March,Arisawa,Hirata,Vars,Vars2,YFZ,HLiu25}
and the spin current\cite{NLSeebeck}. It is notable that both electric field
and temperature gradient are polar vectors, where they change sign under
inversion symmetry $\mathbf{E}\rightarrow -\mathbf{E}$ and $\mathbf{\nabla }%
T\rightarrow -\mathbf{\nabla }T$, while remaining invariant under
time-reversal symmetry. Especially, nonlinear Seebeck effect\cite%
{Zeng,March,Arisawa,Hirata,Vars,Vars2,YFZ,NLSeebeck} and nonlinear Nernst
effect\cite{Yu,Karki,Zeng,Vars,Vars2,NLSeebeck,HLiu25} are such phenomena
that longitudinal and transverse current are induced by a nonlinear response
to a temperature gradient, respectively. The nonlinear Edelstein effect\cite%
{XuEdel,Baek,Trama,Jia} is a phenomenon, where magnetization is induced by a
nonlinear response to electric field. In contrast, magnetization induced by a
nonlinear response to a temperature gradient is yet to be explored.

Altermagnets\cite{SmejX,SmejX2} have emerged as one of the most active
fields in condensed matter physics. Because they possess zero net
magnetization, they are promising candidates for future ultrafast and
ultradense magnetic memories. We investigate the two-dimensional electron
system coupled to an altermagnet. The prominent phenomenon is the emergence
of a distinctive spin-split electron band structure which preserves
inversion symmetry but breaks time-reversal symmetry. In the presence of
inversion symmetry, a linear magnetization response to a temperature
gradient is forbidden, whereas a second-order nonlinear response is allowed.

Odd-parity magnets\cite{Hayami,pwave,He,Luo,APEX} share similarities with
altermagnets in that they also exhibit characteristic spin-split electron
band structures in the coupled electron system, where time-reversal symmetry
is preserved while inversion symmetry is broken  in the coupled
electronic band structure. We note that time-reversal symmetry is preserved
while inversion symmetry is broken in the coupled electronic band structure.
On the other hand, time-reversal symmetry in odd-parity magnets themselves
is broken as a feature common to all magnetic systems.

Altermagnets and odd-parity magnets together form a broader class known as $%
X $-wave magnets, which include $d$-wave, $g$-wave, $i$-wave altermagnets
and $p$-wave, $f$-wave odd-parity magnets.

Spin-split band structures induced by $d$-wave, $g$-wave, $i$-wave
altermagnets are described by $k^{N_{X}}\sin N_{X}\phi $ around the band
bottom, where $N_{X}=2,4$\ and $6$, respectively. On the other hand, there
are $d^{\prime }$-wave, $g^{\prime }$-wave, $i^{\prime }$-wave altermagnets%
\cite{APEX} as well, which are described by $k^{N_{X}}\cos N_{X}\phi $
around the band bottom. $d$-wave altermagnets are also known as $d_{xy}$%
-wave, while $d^{\prime }$-wave altermagnets are also known as $%
d_{x^{2}-y^{2}}$-wave altermagnets.

In this paper, we first derive a general formula for the magnetization
induced by a temperature gradient in the electron system, valid up to
arbitrary orders in nonlinear response,  when the temperature profile is
linear.  We apply this formula to the tight-binding model of the $X$-wave
magnet. Among them, $d^{\prime }$-wave, $g^{\prime }$-wave, $i^{\prime }$%
-wave altermagnets exhibit magnetization induced by the second-order
nonlinear response to a temperature gradient. Using a high-temperature
expansion, we obtain an analytic expression for the induced magnetization.
In contrast, all odd-parity magnets do not exhibit magnetization induced by
temperature gradient owing to the presence of time-reversal symmetry.  We
then present the results when the temperature profile is parabolic
satisfying $\partial _{x}^{2}T\neq 0$, which is realized when the diffusion
constant $\kappa $ has a temperature dependence. In particular, when $\kappa
\propto 1/T$ as observed\cite{Feri} around room temperature in an
altermagnet candidate RuO$_{2}$, the results for the parabolic temperature
profile are simply modified by factor $2$. 

\section{$X$-wave magnets}

We investigate the two-dimensional electron system associated with a
metallic $X$-wave magnet or a metallic thin film attached to an insulating $%
X $-wave magnet. The prominent phenomenon is the emergence of a distinctive
spin-split electron band structure. The simplest model is the two-band
Hamiltonian\cite{GI,Planar,MTJ,APEX,NLSeebeck} given by 
\begin{equation}
H=H_{\text{kine}}\left( \mathbf{k}\right) +Jf_{X}\left( \mathbf{k}\right)
\sigma _{z},  \label{BasicHamil}
\end{equation}%
where the first term represents the kinetic term of electrons, while the
second term represents the\ band splitting described by the function $%
f_{X}\left( \mathbf{k}\right) $\ with the coupling constant $J$\ and the
Pauli matrix $\sigma _{z}$. Around the band bottom, the spin-split function $%
f_{X}\left( \mathbf{k}\right) $ is explicitly given by 
\begin{align}
f_{p}\left( \mathbf{k}\right) & =ak_{x}=ak\cos \phi ,  \label{f2p} \\
f_{d}\left( \mathbf{k}\right) & =2a^{2}k_{x}k_{y}=a^{2}k^{2}\sin 2\phi ,
\label{f2d} \\
f_{f}\left( \mathbf{k}\right) & =a^{3}k_{x}\left(
k_{x}^{2}-3k_{y}^{2}\right) =a^{3}k^{3}\cos 3\phi , \\
f_{g}\left( \mathbf{k}\right) & =4a^{4}k_{x}k_{y}\left(
k_{x}^{2}-k_{y}^{2}\right) =a^{4}k^{4}\sin 4\phi , \\
f_{i}\left( \mathbf{k}\right) & =2a^{6}k_{x}k_{y}\left(
3k_{x}^{2}-k_{y}^{2}\right) \left( k_{x}^{2}-3k_{y}^{2}\right)
=a^{6}k^{6}\sin 6\phi  \label{f2i}
\end{align}%
for the $X$-wave magnet, and%
\begin{align}
f_{p^{\prime }}\left( \mathbf{k}\right) =& ak_{y}=ak\sin \phi ,  \label{f2p2}
\\
f_{d^{\prime }}\left( \mathbf{k}\right) =& a^{2}\left(
k_{x}^{2}-k_{y}^{2}\right) =a^{2}k^{2}\cos 2\phi , \\
f_{f^{\prime }}\left( \mathbf{k}\right) =& a^{3}k_{y}\left(
3k_{x}^{2}-k_{y}^{2}\right) =a^{3}k^{3}\sin 3\phi , \\
f_{g^{\prime }}\left( \mathbf{k}\right) =& a^{4}\left(
k_{x}^{2}-k_{y}^{2}-2k_{x}k_{y}\right) \left(
k_{x}^{2}-k_{y}^{2}+2k_{x}k_{y}\right)  \notag \\
=& a^{4}k^{4}\cos 4\phi , \\
f_{i^{\prime }}\left( \mathbf{k}\right) =& 2a^{6}\left(
k_{x}^{2}-k_{y}^{2}\right) \left(
k_{x}^{4}+k_{y}^{4}-14k_{x}^{2}k_{y}^{2}\right)  \notag \\
=& a^{6}k^{6}\cos 6\phi  \label{f2i2}
\end{align}%
for the $X^{\prime }$-wave magnet, where $k_{x}=k\cos \phi $, $k_{y}=k\sin
\phi $, and $a$ is the lattice constant. We note that the $d$-wave
altermagnet described by the function $f_{d}\left( \mathbf{k}\right) $ is
commonly called the $d_{xy}$-wave altermagnet and $f_{d^{\prime }}\left( 
\mathbf{k}\right) $ is commonly called the $d_{x^{2}-y^{2}}$-wave
altermagnet. The $X$-wave magnet has $N_{X}$ nodes in the band structure,
where $N_{X}=1,2,3,4,6$ for $X=p,d,f,g,i$, respectively.

We make a symmetry analysis of the Hamiltonian (\ref{BasicHamil}).
Time-reversal symmetry is defined by%
\begin{equation}
TH\left( \mathbf{k}\right) T^{-1}=H\left( -\mathbf{k}\right)
\end{equation}%
with the time-reversal symmetry operator $T=i\sigma _{y}K$, where $K$ is the
complex conjugation operator. Inversion symmetry is defined by%
\begin{equation}
PH\left( \mathbf{k}\right) P^{-1}=H\left( -\mathbf{k}\right) .
\end{equation}%
In the $d$-wave, the $g$-wave and the $i$-wave altermagnets, time-reversal
symmetry is broken but inversion symmetry is preserved%
\begin{align}
Tf_{X}\left( \mathbf{k}\right) \sigma _{z}T^{-1} &=-f_{X}\left( -\mathbf{k}%
\right) , \\
Pf_{X}\left( \mathbf{k}\right) \sigma _{z}P^{-1} &=f_{X}\left( -\mathbf{k}%
\right) \sigma _{z}.
\end{align}%
Hence, time-reversal symmetry is broken and inversion symmetry is preserved
in the Hamiltonian $H\left( \mathbf{k}\right) $ for altermagnets. On the
other hand, in the $p$-wave and the $f$-wave odd-parity magnets, inversion
symmetry is broken but time-reversal symmetry is preserved%
\begin{align}
Tf_{X}\left( \mathbf{k}\right) \sigma _{z}T^{-1} &=f_{X}\left( -\mathbf{k}%
\right) \sigma _{z}, \\
Pf_{X}\left( \mathbf{k}\right) \sigma _{z}P^{-1} &=-f_{X}\left( -\mathbf{k}%
\right) \sigma _{z}.
\end{align}%
Hence, time-reversal symmetry is preserved and inversion symmetry is broken
in the Hamiltonian $H\left( \mathbf{k}\right) $ for odd-parity magnets.

Here, we make an important remark. We have introduced the notion of the $X$%
-wave magnet on the basis of the continuum model, where the Hamiltonians (%
\ref{BasicHamil}) for the $X$-wave magnet and the $X^{\prime }$-magnet are
related by the change of the angle $\phi \rightarrow \phi +\pi /(2N_{X})$,
as follows from Eqs.(\ref{f2p})\symbol{126}(\ref{f2i}) and Eqs.(\ref{f2p2})%
\symbol{126}(\ref{f2i2}). However, the continuum model is valid only in low
energy physics. It is necessary to use the tight-binding model to investigate
physics at high temperature, where such a symmetry is lost. In general, the $%
X$-wave magnet and the $X^{\prime }$-magnet are independent objects.

\section{Temperature gradient induced magnetization}

We study a two-dimensional electron system described by the Hamiltonian (\ref%
{BasicHamil}). The linear response to magnetization induced by a temperature
gradient is determined by 
\begin{equation}
M_{\mu }=\chi _{\mu \nu }\nabla _{\nu }T,
\end{equation}%
where $\chi _{\mu \nu }$ is the susceptibility. Electric-field induced
magnetization is prohibited in the inversion symmetric system because the
magnetization is an axial vector, where it does not flip its sign under
inversion symmetry operation but flips its sign under time-reversal symmetry
operation. Next, we consider a second-order nonlinear response associated
with electric-field induced magnetization%
\begin{equation}
M_{\rho }=\chi _{\rho \mu \nu }\left( \nabla _{\mu }T\right) \nabla _{\nu }T,
\end{equation}%
where $\chi _{\rho \mu \nu }$ is the nonlinear susceptibility. Both the left
and right hand sides are invariant under inversion symmetry. Hence, nonzero
magnetization is not prohibited for inversion symmetric systems. On the
other hand, the system must break time-reversal symmetry because the
left-hand side is time-reversal symmetry odd but the right-hand side is
time-reversal symmetry even.

The expectation value of the magnetization is given by%
\begin{equation}
\mathbf{M}=\int \frac{d^{2}k}{\left( 2\pi \right) ^{2}}\mathbf{S}\left( 
\mathbf{k}\right) f\left( \mathbf{k}\right) ,
\end{equation}%
where $\mathbf{S}\left( \mathbf{k}\right) \equiv g\mu _{\text{B}%
}\left\langle \psi \right\vert \mathbf{\sigma }\left\vert \psi \right\rangle
/2$ is the expectation value of the spin, $\mu _{\text{B}}$ is the Bohr
magneton and $g$ is the g factor.

We  first  assume a spatially uniform temperature gradient along the $%
x $ axis such that $\partial _{x}^{2}T=0$.  We discuss the case for $%
\partial _{x}^{2}T\neq 0$ in Sec.VII. 

By using the nonequilibrium Fermi distribution function $f$ determined by
the Boltzmann equation, magnetization induced by the $\ell $-th order
nonlinear response to a temperature gradient is calculated from the formula%
\begin{equation}
\mathbf{M}^{\left( x^{\ell }\right) }=\left( -\frac{\tau }{\hbar }\partial
_{x}T\right) ^{\ell }\int \frac{d^{2}k}{\left( 2\pi \right) ^{2}}\mathbf{S}%
\left( \mathbf{k}\right) \left( \frac{\partial \varepsilon }{\partial k_{x}}%
\right) ^{\ell }\frac{\partial ^{\ell }f^{\left( 0\right) }}{\partial
T^{\ell }},
\end{equation}%
where $\tau $ is the relaxation time, $\varepsilon $ is the energy and 
\begin{equation}
f^{\left( 0\right) }\left( \mathbf{k}\right) =\frac{1}{\exp [\frac{%
\varepsilon \left( \mathbf{k}\right) -\mu }{k_{\text{B}}T\left( \mathbf{r}%
\right) }]+1}
\end{equation}%
is the Fermi distribution function at equilibrium. At high temperature, the
magnetization is approximated as%
\begin{equation}
\mathbf{M}^{\left( x^{\ell }\right) }\simeq -\left( \frac{\tau }{\hbar }%
\partial _{x}T\right) ^{\ell }\frac{\ell !}{4k_{B}T^{\ell +1}}\int \frac{%
d^{2}k}{\left( 2\pi \right) ^{2}}\mathbf{S}\left( \mathbf{k}\right) \left( 
\frac{\partial \varepsilon }{\partial k_{x}}\right) ^{\ell }\left(
\varepsilon -\mu \right) ,  \label{Ml}
\end{equation}%
where we have used%
\begin{equation}
\frac{\partial ^{\ell }f^{\left( 0\right) }}{\partial T^{\ell }}\simeq
\left( -1\right) ^{\ell +1}\frac{\ell !}{4k_{B}T^{\ell +1}}\left(
\varepsilon -\mu \right) .  \label{fex2}
\end{equation}%
See Appendix for derivation of Eq.(\ref{Ml}).

In this system, the spin is diagonal $s=\pm 1$. Hence, the magnetization
formula (\ref{Ml}) is simplified as%
\begin{align}
& M_{z}^{\left( x^{\ell }\right) }  \notag \\
\simeq & \frac{g\mu _{\text{B}}}{2}\Big(-\frac{\tau }{\hbar }\partial _{x}T%
\Big)^{\ell }\frac{\ell !}{4k_{B}T^{\ell +1}}\sum_{s=\pm 1}s\frac{d^{2}k}{%
\left( 2\pi \right) ^{2}}\Big(\frac{\partial \varepsilon _{s}}{\partial k_{x}%
}\Big)^{\ell }\left( \varepsilon _{s}-\mu \right) ,
\end{align}%
where%
\begin{equation}
\varepsilon _{s}=H_{\text{kine}}\left( \mathbf{k}\right) +sJf_{X}\left( 
\mathbf{k}\right)
\end{equation}%
is the energy for spin $s=\pm 1$.

For the second-order nonlinear response, it is explicitly given by%
\begin{align}
M_{z}^{\left( x^{2}\right) }\simeq & -\left( \frac{\tau }{\hbar }\partial
_{x}T\right) ^{2}\frac{g\mu _{\text{B}}J}{2k_{\text{B}}T^{3}}  \notag \\
& \times \frac{d^{2}k}{\left( 2\pi \right) ^{2}}\Big[2(H_{\text{kine}}-\mu
)\partial _{k_{x}}f_{X}\partial _{k_{x}}H_{\text{kine}}  \notag \\
& \hspace{0.6cm}+f_{X}(\partial _{k_{x}}H_{\text{kine}})^{2}+J^{2}f_{X}%
\left( \partial _{k_{x}}f_{X}\right) ^{2}\Big].
\end{align}%
If $f_{X}$ is odd for $k_{x}$, we find $M_{z}^{\left( x^{2}\right) }=0$
because $H_{\text{kine}}$ is even for $k_{x}$. Hence, there are no
second-order nonlinear response in $d$-wave, $g$-wave and $i$-wave
altermagnets. On the other hand, it is nontrivial for $d^{\prime }$-wave, $%
g^{\prime }$-wave and $i^{\prime }$-wave altermagnets. We will analyze
explicitly these altermagnets in the following sections.

\begin{figure}[t]
\centerline{\includegraphics[width=0.48\textwidth]{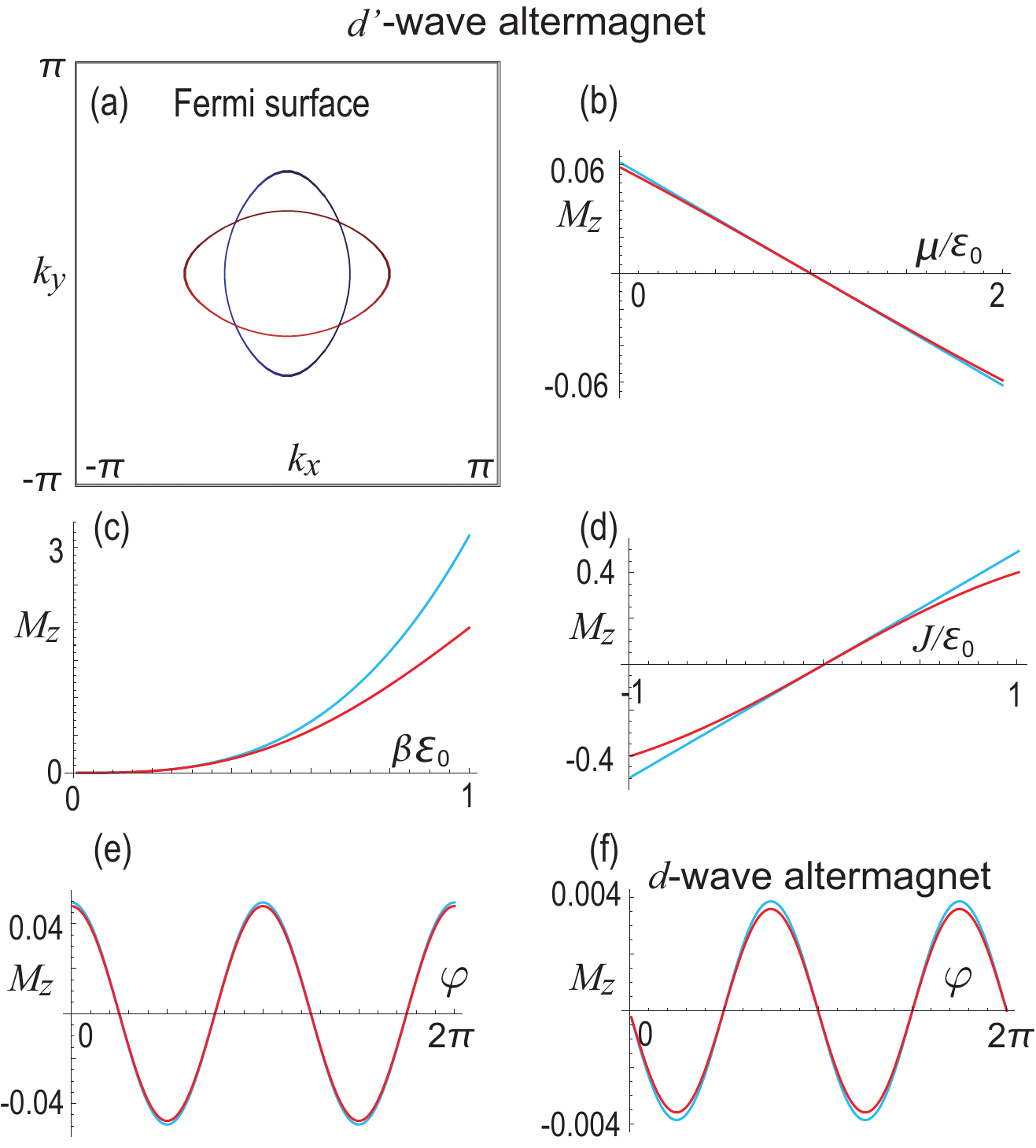}}
\caption{$d^{\prime }$-wave altermagnet. Magnetization induced by
temperature gradient in units of $\frac{g\protect\mu _{\text{B}}}{2\left( 2%
\protect\pi \right) ^{2}}\left( -\frac{\protect\tau }{\hbar }\partial
_{x}T\right) ^{2}.$ (a) Fermi surfaces. Red oval represents up spin, while
blue oval represents down spin. (b) $\protect\mu $ dependence. We have set $%
J/\protect\varepsilon _{0}=0.1$ and $\protect\beta \protect\varepsilon %
_{0}=1/4$. (c) $\protect\beta $ dependence. We have set $J/\protect%
\varepsilon _{0}=0.1$ and $\protect\mu /\protect\varepsilon _{0}=0.2$. (d) $%
J $ dependence. Magnetization as a function of $\protect\varphi $\ in units
of $\frac{g\protect\mu _{\text{B}}}{2}\left( -\frac{\protect\tau }{\hbar }%
\partial _{X}T\right) ^{2}$ of (e) $d^{\prime }$-wave altermagnet and (f) $d$%
-wave altermagnet. We have set $\protect\beta \protect\varepsilon _{0}=1/4$
and $\protect\mu /\protect\varepsilon _{0}=0.2$. Red curves represent
numerical results, while cyan curves represent analytic results based on
high-temperature expansion. We have set $\protect\varepsilon _{0}=4\hbar
^{2}/\left( ma^{2}\right) $.}
\label{FigD}
\end{figure}

We also study magnetization induced by the temperature gradient with an
arbitrary angle $\varphi $, where the new coordinate is defined by%
\begin{equation}
X\equiv x\cos \varphi -y\sin \varphi ,\quad Y\equiv x\sin \varphi +y\cos
\varphi .
\end{equation}%
The magnetization is calculated by changing $\partial _{x}$ to $\partial
_{X} $ with%
\begin{equation}
\partial _{X}=\frac{\partial x}{\partial X}\frac{\partial }{\partial x}+%
\frac{\partial y}{\partial X}\frac{\partial }{\partial y}=\cos \varphi \frac{%
\partial }{\partial x}-\sin \varphi \frac{\partial }{\partial y}
\end{equation}%
in Eq.(\ref{Ml}) together with%
\begin{equation}
\partial _{k_{X}}=\frac{\partial k_{x}}{\partial k_{X}}\frac{\partial }{%
\partial k_{x}}+\frac{\partial k_{y}}{\partial k_{X}}\frac{\partial }{%
\partial k_{y}}=\cos \varphi \frac{\partial }{\partial k_{x}}-\sin \varphi 
\frac{\partial }{\partial k_{y}}.
\end{equation}%
There is a relation%
\begin{equation}
M_{z}^{\left( X^{2}\right) }=M_{z}^{\left( x^{2}\right) }\cos ^{2}\varphi
+M_{z}^{\left( y^{2}\right) }\sin ^{2}\varphi -M_{z}^{\left( xy\right) }\sin
2\varphi .
\end{equation}

\section{$d^{\prime }$-wave altermagnet}

The tight-binding model (\ref{BasicHamil}) for the $d^{\prime }$-wave
altermagnet is given by\cite%
{SmejRev,SmejX,SmejX2,Zu2023,Gho,Li2023,EzawaAlter,EzawaMetricC}%
\begin{equation}
H=H_{\text{kine,sq}}+Jf_{d^{\prime }}\sigma _{z}
\end{equation}
with%
\begin{equation}
H_{\text{kine,sq}}=\frac{\hbar ^{2}}{ma^{2}}\left( 2-\cos ak_{x}-\cos
ak_{y}\right) ,  \label{Hsq}
\end{equation}%
and%
\begin{equation}
f_{d^{\prime }}=2\left( \cos ak_{y}-\cos ak_{x}\right) ,
\end{equation}%
where $a$ is the lattice constant. In the following, we set $\overline{m}%
\equiv ma^{2}/\hbar ^{2}$. The Fermi surface is shown in Fig.\ref{FigD}(a).

The magnetization is analytically obtained from Eq.(\ref{Ml}) as%
\begin{equation}
\frac{M_{z}^{\left( x^{2}\right) }}{\frac{g\mu _{\text{B}}}{2\left( 2\pi
\right) ^{2}}\left( \frac{\tau }{\hbar }\partial _{x}T\right) ^{2}}\simeq -%
\frac{8\pi ^{2}J\left( \overline{m}\mu -2\right) }{\overline{m}^{2}k_{\text{B%
}}T^{3}}.  \label{EqD}
\end{equation}%
The magnetization is shown as a function of $\mu $\ in Fig.\ref{FigD}(b), as
a function of $\beta $\ in Fig.\ref{FigD}(c) and as a function of $J$\ in
Fig.\ref{FigD}(d). The analytical results (cyan curves) obtained by using
high-temperature expansion well agree with the numerical results (red
curves) without using the expansion.

The angle dependence is analytically obtained as

\begin{align}
\frac{M_{z}^{\left( y^{2}\right) }}{\frac{g\mu _{\text{B}}}{2\left( 2\pi
\right) ^{2}}\left( \frac{\tau }{\hbar }\partial _{y}T\right) ^{2}}& =-\frac{%
M_{z}^{\left( x^{2}\right) }}{\frac{g\mu _{\text{B}}}{2\left( 2\pi \right)
^{2}}\left( \frac{\tau }{\hbar }\partial _{x}T\right) ^{2}}, \\
\frac{M_{z}^{\left( xy\right) }}{\frac{g\mu _{\text{B}}}{2\left( 2\pi
\right) ^{2}}\left( \frac{\tau }{\hbar }\right) ^{2}\partial _{x}T\partial
_{y}T}& =0,
\end{align}%
and%
\begin{equation}
\frac{M_{z}^{\left( X^{2}\right) }}{\frac{g\mu _{\text{B}}}{2\left( 2\pi
\right) ^{2}}\left( \frac{\tau }{\hbar }\partial _{X}T\right) ^{2}}\simeq 
\frac{8\pi ^{2}J\left( \overline{m}\mu -2\right) }{\overline{m}^{2}k_{\text{B%
}}T^{3}}\cos 2\varphi .
\end{equation}%
The magnetization is shown as a function of $\mu $\ in Fig.\ref{FigD}(e).

We also study the $d$-wave altermagnet, where 
\begin{equation}
f_{d}=\sin ak_{x}\sin ak_{y}.
\end{equation}%
The angle dependence of the magnetization is analytically obtained as%
\begin{align}
\frac{M_{z}^{\left( x^{2}\right) }}{\frac{g\mu _{\text{B}}}{2\left( 2\pi
\right) ^{2}}\left( \frac{\tau }{\hbar }\partial _{x}T\right) ^{2}} &=-\frac{%
M_{z}^{\left( y^{2}\right) }}{\frac{g\mu _{\text{B}}}{2\left( 2\pi \right)
^{2}}\left( \frac{\tau }{\hbar }\partial _{y}T\right) ^{2}}=0, \\
\frac{M_{z}^{\left( xy\right) }}{\frac{g\mu _{\text{B}}}{2\left( 2\pi
\right) ^{2}}\left( \frac{\tau }{\hbar }\right) ^{2}\partial _{x}T\partial
_{y}T} &\simeq -\frac{\pi ^{2}J}{m^{2}k_{\text{B}}T^{3}}.
\end{align}%
The angle dependence is obtained as%
\begin{equation}
\frac{M_{z}^{\left( X^{2}\right) }}{\frac{g\mu _{\text{B}}}{2\left( 2\pi
\right) ^{2}}\left( \frac{\tau }{\hbar }\partial _{X}T\right) ^{2}}\simeq -%
\frac{\pi ^{2}J}{m^{2}k_{\text{B}}T^{3}}\sin 2\varphi
\end{equation}%
up to any order in $J$. The magnetization is shown as a function of $\mu $\
in Fig.\ref{FigD}(f). The angle dependence between (e) and (f) is different
by $\pi /4$, which coincides with the angle difference between the $%
d^{\prime }$-wave and $d$-wave altermagnets. 
\begin{figure}[t]
\centerline{\includegraphics[width=0.48\textwidth]{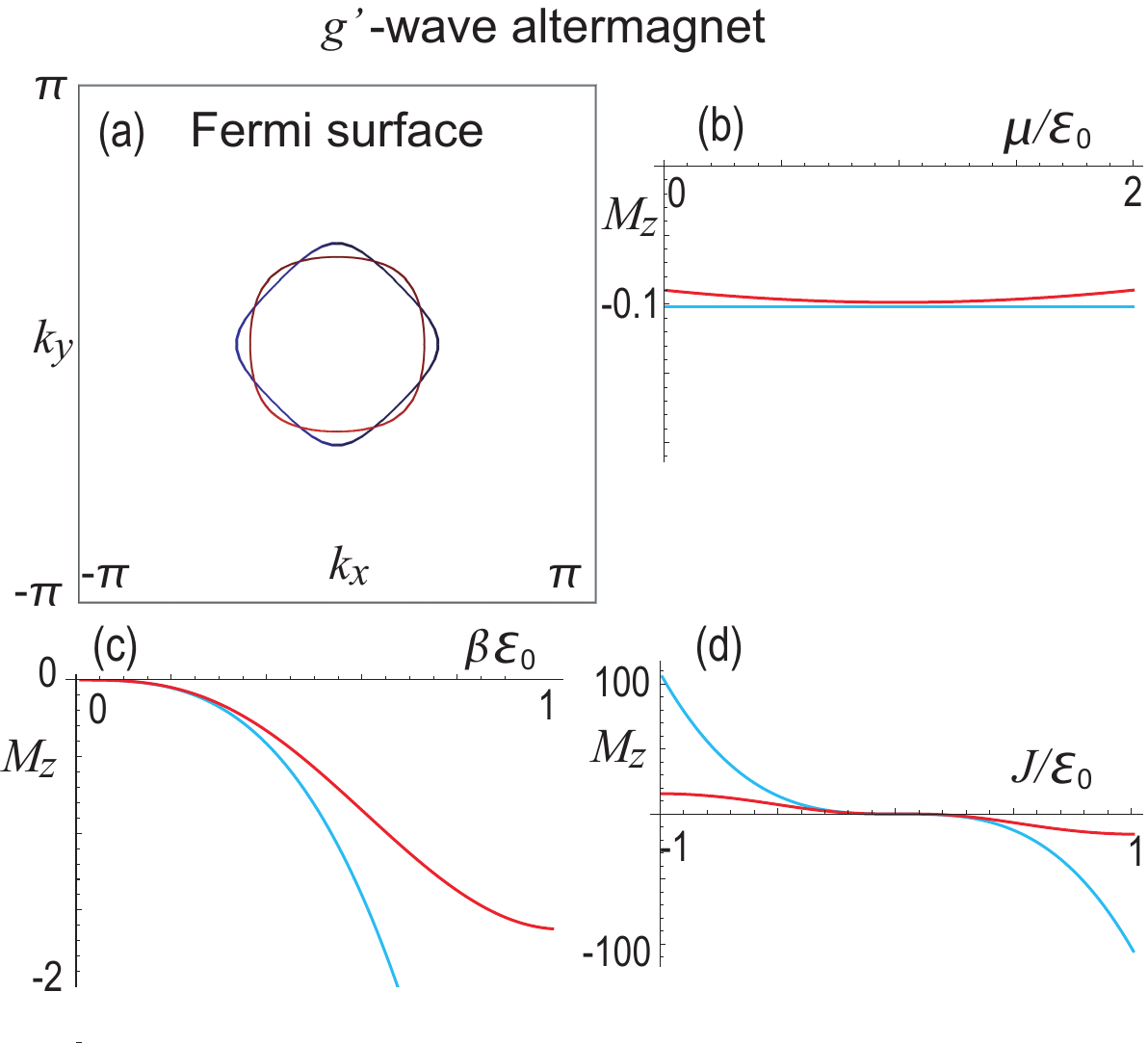}}
\caption{$g^{\prime }$-wave altermagnet. Magnetization induced by
temperature gradient in units of $\frac{g\protect\mu _{\text{B}}}{2\left( 2%
\protect\pi \right) ^{2}}\left( -\frac{\protect\tau }{\hbar }\partial
_{x}T\right) ^{2}.$ (a) Fermi surface. (b) $\protect\mu $ dependence. We
have set $J/\protect\varepsilon _{0}=0.1$ and $\protect\beta \protect%
\varepsilon _{0}=1/4$. (c) $\protect\beta $ dependence. We have set $J/%
\protect\varepsilon _{0}=0.1$ and $\protect\mu /\protect\varepsilon _{0}=0.2$%
. (d) $J$ dependence. We have set $\protect\beta \protect\varepsilon %
_{0}=1/4 $ and $\protect\mu /\protect\varepsilon _{0}=0.2$. Red curves
represent numerical results, while cyan curves represent analytic results
based on high-temperature expansion. We have set $\protect\varepsilon %
_{0}=4\hbar ^{2}/\left( ma^{2}\right) $. }
\label{FigG}
\end{figure}

\section{$g^{\prime }$-wave altermagnets}

The tight-binding model (\ref{BasicHamil}) for the $g^{\prime }$-wave
altermagnet is%
\begin{equation}
H=H_{\text{kine,sq}}+Jf_{g^{\prime }}\sigma _{z}
\end{equation}
with Eq.(\ref{Hsq}) and%
\begin{align}
f_{g^{\prime }}=& 4\left( \cos ak_{y}-\cos ak_{x}-\sin ak_{x}\sin
ak_{y}\right)  \notag \\
& \times \left( \cos ak_{y}-\cos ak_{x}+\sin ak_{x}\sin ak_{y}\right) .
\end{align}
The Fermi surface is shown in Fig.\ref{FigG}(a).

The magnetization is analytically obtained as%
\begin{equation}
\frac{M_{z}^{\left( x^{2}\right) }}{\frac{g\mu _{\text{B}}}{2\left( 2\pi
\right) ^{2}}\left( \frac{\tau }{\hbar }\partial _{x}T\right) ^{2}}\simeq -%
\frac{\pi ^{2}J\left( 1-684\overline{m}^{2}J^{2}\right) }{\overline{m}^{2}k_{%
\text{B}}T^{3}}  \label{EqG}
\end{equation}%
up to any order in $J$. The magnetization is shown as a function of $\mu $\
in Fig.\ref{FigG}(b), as a function of $\beta $\ in Fig.\ref{FigG}(c) and as
a function of $J$\ in Fig.\ref{FigG}(d). The analytical result obtained by
using high-temperature expansion well agrees with the numerical result
without using the expansion. 
\begin{figure}[t]
\centerline{\includegraphics[width=0.48\textwidth]{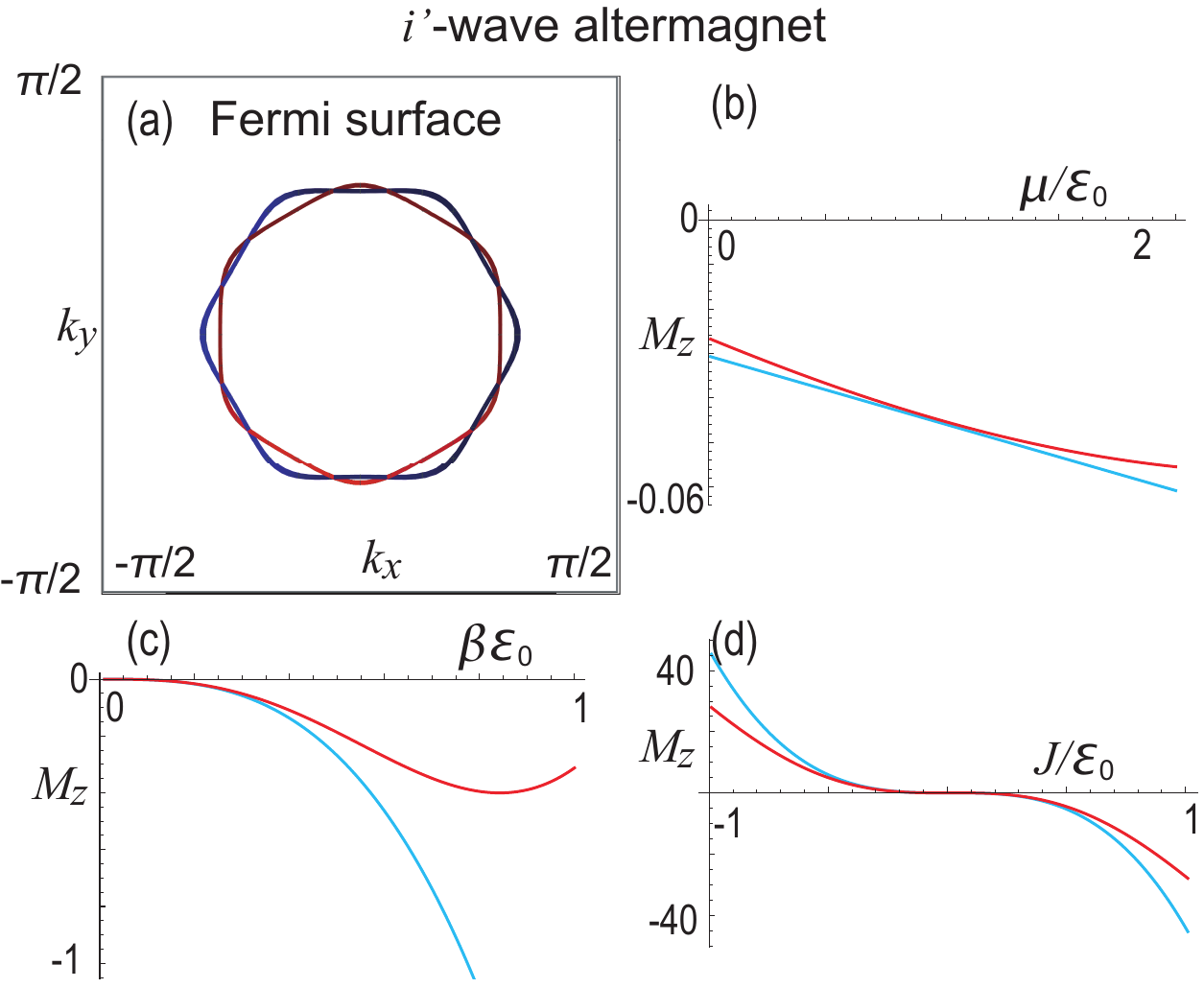}}
\caption{$i^{\prime }$-wave altermagnet. Magnetization induced by
temperature gradient in units of $\frac{g\protect\mu _{\text{B}}}{2}\left( -%
\frac{\protect\tau }{\hbar }\partial _{x}T\right) ^{2}.$ (a) Fermi surface.
(b) $\protect\mu $ dependence. We have set $J/\protect\varepsilon _{0}=0.1$
and $\protect\beta \protect\varepsilon _{0}=1/4$. (c) $\protect\beta $
dependence. We have set $J/\protect\varepsilon _{0}=0.1$ and $\protect\mu /%
\protect\varepsilon _{0}=0.2$. (d) $J$ dependence. We have set $\protect%
\beta \protect\varepsilon _{0}=1/4$ and $\protect\mu /\protect\varepsilon %
_{0}=0.2$. Red curves represent numerical results, while cyan curves
represent analytic results based on high-temperature expansion. We have set $%
\protect\varepsilon _{0}=4\hbar ^{2}/\left( ma^{2}\right) $. }
\label{FigI}
\end{figure}
We find%
\begin{equation}
M_{z}^{\left( y^{2}\right) }=M_{z}^{\left( x^{2}\right) },\qquad
M_{z}^{\left( xy\right) }=0
\end{equation}%
Hence, there is no angle dependence for $g^{\prime }$-wave altermagnets.

We also study the $g$-wave altermagnet with%
\begin{equation}
f_{g}=8\sin ak_{x}\sin ak_{y}\left( \cos ak_{y}-\cos ak_{x}\right) .
\end{equation}%
We find%
\begin{equation}
M_{z}^{\left( x^{2}\right) }=M_{z}^{\left( y^{2}\right) }=M_{z}^{\left(
xy\right) }=0,
\end{equation}%
because the system is odd for the mirror operations. Hence, the
magnetization is zero for $g$-wave altermagnet for an arbitrary angle $%
\varphi $.

\section{$i^{\prime }$-wave altermagnets}

The tight-binding model (\ref{BasicHamil}) for the $i^{\prime }$-wave
altermagnet is%
\begin{equation}
H=H_{\text{kine,tri}}+Jf_{i^{\prime }}\sigma _{z}
\end{equation}
with the kinetic term%
\begin{equation}
H_{\text{kine,tri}}=\frac{-2\hbar ^{2}}{3ma^{2}}\Big(\sum_{j=0}^{2}\cos
\left( a\mathbf{n}_{j}^{\text{A}}\cdot \mathbf{k}\right) -3\Big),
\end{equation}%
and%
\begin{align}
f_{i^{\prime }}=& 16\Bigg[\Big(\prod\limits_{j=0}^{2}\sin (a\mathbf{n}_{j}^{%
\text{A}}\cdot \mathbf{k})\Big)^{2}  \notag \\
& \hspace{0.6cm}-\Big(\frac{1}{3\sqrt{3}}\prod\limits_{j=0}^{2}\sin (\sqrt{3}%
a\mathbf{n}_{j}^{\text{B}}\cdot \mathbf{k})\Big)^{2}\Bigg],
\end{align}%
where we have defined%
\begin{align}
\mathbf{n}_{j}^{\text{A}}& =\left( \cos \frac{2\pi j}{3},\sin \frac{2\pi j}{3%
}\right) ,  \notag \\
\mathbf{n}_{j}^{\text{B}}& =\left( \sin \frac{2\pi j}{3},\cos \frac{2\pi j}{3%
}\right)
\end{align}%
with $j=0,1,2$. \ The Fermi surface is shown in Fig.\ref{FigI}(a).

The magnetization is analytically obtained as%
\begin{equation}
\frac{M_{z}^{\left( x^{2}\right) }}{\frac{g\mu _{\text{B}}}{2\left( 2\pi
\right) ^{2}}\left( \frac{\tau }{\hbar }\partial _{x}T\right) ^{2}}\simeq -%
\frac{4\pi ^{2}J}{3^{3}\sqrt{3}\overline{m}^{2}k_{\text{B}}T^{3}}-\frac{%
372058\pi ^{2}J^{3}}{3^{7}\sqrt{3}k_{\text{B}}T^{3}}  \label{EqI}
\end{equation}%
up to any order in $J$. We find%
\begin{equation}
M_{z}^{\left( y^{2}\right) }=M_{z}^{\left( x^{2}\right) },\qquad
M_{z}^{\left( xy\right) }=0
\end{equation}%
Hence, there is no angle dependence for $i^{\prime }$-wave altermagnets. In
addition, the magnetization is zero for $i^{\prime }$-wave altermagnet for
an arbitrary angle $\varphi $.

The magnetization is shown as a function of $\mu $\ in Fig.\ref{FigI}(b), as
a function of $\beta $\ in Fig.\ref{FigI}(c) and as a function of $J$\ in
Fig.\ref{FigI}(d). The analytical result obtained by using high-temperature
expansion well agree with the numerical result without using the expansion.

We also study the $i$-wave altermagnet, where%
\begin{equation}
f_{i}=16J\sigma _{z}\sin ak_{z}\prod\limits_{j=0}^{2}\sin \left( a\mathbf{n}%
_{j}^{\text{A}}\cdot \mathbf{k}\right) \prod\limits_{j=0}^{2}\sin \left( 
\sqrt{3}a\mathbf{n}_{j}^{\text{B}}\cdot \mathbf{k}\right) .
\end{equation}%
We find%
\begin{equation}
M_{z}^{\left( x^{2}\right) }=M_{z}^{\left( y^{2}\right) }=M_{z}^{\left(
xy\right) }=0,
\end{equation}%
because the system is odd for the mirror operations. Hence, the
magnetization is zero for $i$-wave altermagnet for an arbitrary angle $%
\varphi $.

\section{Parabolic temperature profile}

 We have so far assumed that $\partial _{x}^{2}T=0$, which is not valid
when the diffusion coefficient $\kappa $ has a temperature dependence. The
heat equation under steady-state conditions reads%
\begin{equation}
\nabla \left( \kappa \nabla T\right) =0,
\end{equation}%
where $\kappa $ is the diffusion coefficient. We have $\nabla ^{2}T=0$ only
when $\kappa $ is a constant. Otherwise, we have%
\begin{equation}
\frac{\partial ^{2}T}{\partial x^{2}}=-\frac{1}{\kappa }\frac{\partial
\kappa }{\partial T}\left( \frac{\partial T}{\partial x}\right) ^{2}.
\label{Tk}
\end{equation}%
Consequently, $\partial _{x}^{2}T\neq 0$ if $\partial _{x}T\neq 0$ when $%
\kappa $ is not a constant. Indeed, it is pointed out\cite{Feri} that%
\begin{equation}
\kappa =c/T  \label{FeriKappa}
\end{equation}%
with a constant $c$ in RuO$_{2}$, which is a candidate material for the $d$%
-wave altermagnet.

When $\partial _{x}^{2}T\neq 0$ in general, the formula for the second-order
response of the magnetization is modified as%
\begin{align}
M_{z}^{\left( x^{2}\right) }=& -g\mu _{\text{B}}\left( \frac{\tau }{\hbar }%
\right) ^{2}\int d^{3}k\mathbf{S}\left( \mathbf{k}\right) \left( \frac{%
\partial \varepsilon }{\partial k_{x}}\right) ^{2}\frac{\partial
^{2}f^{\left( 0\right) }}{\partial x^{2}}  \notag \\
=& -g\mu _{\text{B}}\left( \frac{\tau }{\hbar }\right) ^{2}\int d^{3}k%
\mathbf{S}\left( \mathbf{k}\right) \left( \frac{\partial \varepsilon }{%
\partial k_{x}}\right) ^{2}  \notag \\
& \times \left[ \frac{\partial ^{2}f^{\left( 0\right) }}{\partial T^{2}}%
\left( \partial _{x}T\right) ^{2}+\frac{\partial f^{\left( 0\right) }}{%
\partial T}\partial _{x}^{2}T\right] ,
\end{align}%
where we have used the relation%
\begin{eqnarray}
\frac{\partial ^{2}f^{\left( 0\right) }}{\partial x^{2}} &=&\frac{\partial }{%
\partial x}\left( \frac{\partial T}{\partial x}\frac{\partial f^{\left(
0\right) }}{\partial T}\right) =\frac{\partial ^{2}T}{\partial x^{2}}\frac{%
\partial f^{\left( 0\right) }}{\partial T}+\frac{\partial T}{\partial x}%
\frac{\partial T}{\partial x}\frac{\partial }{\partial T}\frac{\partial
f^{\left( 0\right) }}{\partial T}  \notag \\
&=&\frac{\partial f^{\left( 0\right) }}{\partial T}\partial _{x}^{2}T+\frac{%
\partial ^{2}f^{\left( 0\right) }}{\partial T^{2}}\left( \partial
_{x}T\right) ^{2},  \label{paraM}
\end{eqnarray}%
and Eq.(\ref{fl2}) in Appendix. By using Eq.(\ref{fex2}) with $\ell =2$, we
obtain%
\begin{align}
\mathbf{M}^{\left( x^{\ell }\right) }\simeq & \frac{g\mu _{\text{B}}}{2}%
\left( \frac{\tau }{\hbar }\right) ^{2}\left[ \left( \partial _{x}T\right)
^{2}\frac{1}{2k_{B}T^{3}}-\frac{1}{4k_{B}T^{2}}\partial _{x}^{2}T\right] 
\notag \\
& \times \int d^{3}k\mathbf{S}\left( \mathbf{k}\right) \left( \frac{\partial
\varepsilon }{\partial k_{x}}\right) ^{\ell }\left( \varepsilon -\mu \right)
.
\end{align}%
Hence, the second-order response of the magnetization for $\partial
_{x}^{2}T\neq 0$ is given by the following replacement%
\begin{eqnarray}
&&\left( \frac{\tau }{\hbar }\partial _{x}T\right) ^{2}\frac{1}{2k_{B}T^{3}}
\notag \\
&\mapsto &\frac{1}{2k_{B}T^{3}}\left( \frac{\tau }{\hbar }\right) ^{2}\left(
\partial _{x}T\right) ^{2}-\left( \frac{\tau }{\hbar }\right) ^{2}\frac{1}{%
4k_{B}T^{2}}\partial _{x}^{2}T.
\end{eqnarray}%
Namely, Eqs.(\ref{EqD}), (\ref{EqG}) and (\ref{EqI}) are replaced from%
\begin{equation}
\frac{M_{z}^{\left( x^{2}\right) }}{\frac{g\mu _{\text{B}}}{2\left( 2\pi
\right) ^{2}}\left( \frac{\tau }{\hbar }\partial _{x}T\right) ^{2}}=F
\end{equation}%
to%
\begin{equation}
\frac{M_{z}^{\left( x^{2}\right) }}{\frac{g\mu _{\text{B}}}{2\left( 2\pi
\right) ^{2}}\left( \frac{\tau }{\hbar }\right) ^{2}\left[ \frac{1}{%
2k_{B}T^{3}}\left( \partial _{x}T\right) ^{2}-\frac{1}{4k_{B}T^{2}}\partial
_{x}^{2}T\right] }=F.
\end{equation}%
With the use of Eq.(\ref{Tk}), we have%
\begin{eqnarray}
&&\frac{1}{2k_{B}T^{3}}\left( \partial _{x}T\right) ^{2}-\frac{1}{4k_{B}T^{2}%
}\partial _{x}^{2}T  \notag \\
&=&\frac{1}{2k_{B}T^{3}}\left( \partial _{x}T\right) ^{2}+\frac{1}{%
4k_{B}T^{2}}\frac{1}{\kappa }\frac{\partial \kappa }{\partial T}\left( \frac{%
\partial T}{\partial x}\right) ^{2}  \notag \\
&=&\frac{1}{2k_{B}T^{3}}\left( \partial _{x}T\right) ^{2}\left[ 1+\frac{T}{2}%
\frac{1}{\kappa }\frac{\partial \kappa }{\partial T}\right] .
\end{eqnarray}%
For instance, in the case of Eq.(\ref{FeriKappa}), we obtain 
\begin{eqnarray}
&&\frac{1}{2k_{B}T^{3}}\left( \partial _{x}T\right) ^{2}\left[ 1+\frac{T}{2}%
\frac{1}{\kappa }\frac{\partial \kappa }{\partial T}\right]  \notag \\
&=&\frac{1}{2k_{B}T^{3}}\left( \partial _{x}T\right) ^{2}\left[ 1+\frac{T}{2}%
\frac{T}{c}\left( -\frac{c}{T^{2}}\right) \right]  \notag \\
&=&\frac{1}{4k_{B}T^{3}}\left( \partial _{x}T\right) ^{2}.
\end{eqnarray}%
Consequently, the results are modified only by factor 2.

\section{Discussion}

We have investigated the emergence of the magnetization induced by the
second-order response to a temperature gradient in the $X$-wave and $%
X^{\prime }$-wave magnets with $X=d$, $g$ and $i$. It is noteworthy that
there are nontrivial responses in $d^{\prime }$-wave, $g^{\prime }$-wave and 
$i^{\prime }$-wave altermagnets when the temperature gradient is along the $%
x $ or $y$ direction. In addition, there are nontrivial responses in $d$%
-wave altermagnets when the temperature gradient is along an arbitrary
direction. On the other hand, there is no such responses occur in $g$-wave
and $i$-wave altermagnets. It is understood in terms of the mirror symmetry
with respect to the $x$ and $y$ axes. We also note that $g^{\prime }$-wave
and $i^{\prime }$-wave altermagnets have been scarcely studied, despite
being predicted on the basis of symmetry analysis\cite{APEX}.

Magnetization induced by temperature gradient is experimentally observed in
Au\cite{Hou}. We estimate the magnetization induced by the second-order
nonlinear response to the temperature gradient. The magnetization per volume
is estimated as%
\begin{equation}
\frac{g\mu _{\text{B}}}{2a^{3}}\left( \tau v_{\text{F}}\frac{\partial _{x}T}{%
T}\right) ^{2}\frac{\varepsilon -\mu }{k_{\text{B}}T}\sim 0.15\text{A/m},
\end{equation}%
where we have used a typical relaxation time $\tau =3\times 10^{-12}$s, the
Fermi velocity $v_{\text{F}}=10^{6}$m/s, $\partial _{x}T=1$K/mm, $T=300$K, $%
g\mu _{\text{B}}=2\times 10^{-23}$Am$^{2}$, $a=3$\AA , $k_{\text{B}}T=25$meV
and $\varepsilon -\mu =10$meV. By assuming the sample with the cubic whose
length is 1mm, the magnetization is estimated as%
\begin{equation}
0.15\text{A/m}\times \left( 10^{-3}\text{m}\right) ^{3}=1.5\times 10^{-14}%
\text{Am}^{2}.
\end{equation}%
It is smaller than a typical order of the magnetization by the SQUID of the
order of $10^{-11}$Am$^{2}$\cite{Buch}, but is experimentally detectable
because the minimum measurable magnetization by the SQUID is of the order of 
$10^{-14}$Am$^{2}$\cite{Ferra,Hass}. The spin splitting in the band
structure for the $d^{\prime }$-wave altermagnet is 1.4eV in RuO$_{2}$\cite%
{ZLin}. On the other hand, $g$-wave and $i$-wave altermagnets are
theoretically proposed to be realized in van der Waals ferromagnets\cite%
{YLiu}, where the spin splitting is estimated be around 10meV. Then, the
magnetization may be ranging from $10^{-15}$Am$^{2}$ to $10^{-13}$Am$^{2}$.

This work is supported by Grants-in-Aid for Scientific Research from MEXT
KAKENHI (Grant No. 23H00171).

\appendix

\section{Temperature Gradient Induced Magnetization}

We derive the formula for magnetization induced by temperature gradient $%
\mathbf{\nabla }T$ based on the Boltzmann equation. By assuming the spatial
dependence of temperature $T\left( \mathbf{r}\right) $, the Fermi
distribution function is given by 
\begin{equation}
f(\mathbf{r},\mathbf{k})=\frac{1}{\exp [\frac{\varepsilon \left( \mathbf{k}%
\right) -\mu }{k_{\text{B}}T\left( \mathbf{r}\right) }]+1}.
\end{equation}%
The Boltzmann equation is given by%
\begin{equation}
\frac{df(\mathbf{r},\mathbf{k})}{dt}=-\frac{f(\mathbf{r},\mathbf{k}%
)-f^{\left( 0\right) }}{\tau },
\end{equation}%
where $\tau $ is the relaxation time. The total derivative is expanded as%
\begin{align}
\frac{df}{dt}& =\frac{\partial f}{\partial t}+\frac{\partial \mathbf{k}}{%
\partial t}\cdot \frac{\partial f}{\partial \mathbf{k}}+\frac{\partial 
\mathbf{r}}{\partial t}\cdot \frac{\partial f}{\partial \mathbf{r}}  \notag
\\
& =\frac{\partial \mathbf{k}}{\partial t}\cdot \left( \frac{\partial
\varepsilon \left( \mathbf{k}\right) }{\partial \mathbf{k}}\frac{\partial f}{%
\partial \varepsilon }\right) +\frac{\partial \mathbf{r}}{\partial t}\cdot
\left( \frac{\partial T\left( \mathbf{r}\right) }{\partial \mathbf{r}}\frac{%
\partial f}{\partial T}\right) .
\end{align}%
We use the kinetic equations%
\begin{equation}
\frac{\partial \mathbf{r}}{\partial t}=\frac{1}{\hbar }\frac{\partial
\varepsilon \left( \mathbf{k}\right) }{\partial \mathbf{k}},\quad \frac{%
\partial \mathbf{k}}{\partial t}=0
\end{equation}%
to obtain%
\begin{equation}
\frac{1}{\hbar }\left( \frac{\partial \varepsilon \left( \mathbf{k}\right) }{%
\partial \mathbf{k}}\cdot \frac{\partial T\left( \mathbf{r}\right) }{%
\partial \mathbf{r}}\right) \frac{\partial f}{\partial T}=-\frac{f-f^{\left(
0\right) }}{\tau }.
\end{equation}%
Assuming that the temperature gradient is along the $x$ direction, we
measure the induced current along the $b$ direction.  First, we consider
the case $\partial _{x}^{2}T=0$. Next, we consider the case $\partial
_{x}^{2}T\neq 0$. 

First, when $\partial _{x}^{2}T=0$, we expand the Fermi distribution in
powers of $\partial _{x}T$, $f=f^{\left( 0\right) }+f^{\left( 1\right)
}+\cdots $, and we obtain 
\begin{equation}
\frac{1}{\hbar }\left( \frac{\partial \varepsilon \left( \mathbf{k}\right) }{%
\partial \mathbf{k}}\cdot \frac{\partial T\left( \mathbf{r}\right) }{%
\partial \mathbf{r}}\right) \frac{\partial f^{\left( \ell -1\right) }}{%
\partial T}=-\frac{f^{\left( \ell \right) }}{\tau }.
\end{equation}%
It is straightforward to solve the Boltzmann equation recursively and derive
the formula%
\begin{equation}
f^{\left( \ell \right) }=\left( -\frac{\tau }{\hbar }\partial _{x}T\right)
^{\ell }\left( \frac{\partial \varepsilon }{\partial k_{b}}\right) ^{\ell }%
\frac{\partial ^{\ell }f^{\left( 0\right) }}{\partial T^{\ell }},
\end{equation}%
where we have assumed that temperature gradient is along the $x$-axis.

At high temperature $T$, the derivative of the Fermi-distribution function
at equilibrium is approximated as 
\begin{equation}
\frac{\partial ^{\ell }f^{\left( 0\right) }}{\partial T^{\ell }}=\left(
-1\right) ^{\ell +1}\frac{\ell !}{4k_{B}T^{\ell +1}}\left( \varepsilon -\mu
\right) .  \label{fex}
\end{equation}%
By inserting it to the Fermi distribution function at nonequilibrium, we
obtain 
\begin{equation}
f^{\left( \ell \right) }=-\left( \frac{\tau }{\hbar }\partial _{x}T\right)
^{\ell }\left( \frac{\partial \varepsilon }{\partial k_{b}}\right) ^{\ell }%
\frac{\ell !}{4k_{B}T^{\ell +1}}\left( \varepsilon -\mu \right) .
\end{equation}

The expectation value of the magnetization is given by%
\begin{equation}
\mathbf{M}=\int d^{3}k\mathbf{S}\left( \mathbf{k}\right) f,
\end{equation}%
where 
\begin{equation}
\mathbf{S}\left( \mathbf{k}\right) \equiv \frac{g\mu _{\text{B}}}{2}%
\left\langle \psi \right\vert \mathbf{\sigma }\left\vert \psi \right\rangle 
\end{equation}%
is the expectation value of the spin and $\mu _{\text{B}}$ is the Bohr
magneton. By using the Fermi distribution function, magnetization induced by
the $\ell $-th order nonlinear response to a temperature gradient is
calculated from the formula 
\begin{align}
\mathbf{M}^{\left( x^{\ell };b\right) }=& -g\mu _{\text{B}}\left( \frac{\tau 
}{\hbar }\partial _{x}T\right) ^{\ell }\int d^{3}k\mathbf{S}\left( \mathbf{k}%
\right) \left( \frac{\partial \varepsilon }{\partial k_{b}}\right) ^{\ell }%
\frac{\partial ^{\ell }f^{\left( 0\right) }}{\partial T^{\ell }}  \notag \\
\simeq & -\frac{g\mu _{\text{B}}}{2}\left( \frac{\tau }{\hbar }\partial
_{x}T\right) ^{\ell }\frac{\ell !}{4k_{B}T^{\ell +1}}  \notag \\
& \times \int d^{3}k\mathbf{S}\left( \mathbf{k}\right) \left( \frac{\partial
\varepsilon }{\partial k_{b}}\right) ^{\ell }\left( \varepsilon -\mu \right)
.
\end{align}

In the following, we consider the two-band Hamiltonian for a metallic $X$%
-wave magnet,%
\begin{equation}
H=H_{\text{kine}}\left( \mathbf{k}\right) +Jf_{X}\left( \mathbf{k}\right)
\sigma _{z},
\end{equation}%
where $H_{\text{kine}}\left( \mathbf{k}\right) $ is the kinetic term and $%
f_{X}\left( \mathbf{k}\right) $ is the spin-split function characteristic to
the $X$-wave magnet. In this system, the spin is diagonal $\mathbf{S}\left( 
\mathbf{k}\right) =\pm 1$. Hence, the magnetization formula is simplified as 
\begin{align}
M_{z}^{\left( x^{\ell };b\right) }=& -\frac{g\mu _{\text{B}}}{2}\left( \frac{%
\tau }{\hbar }\partial _{x}T\right) ^{\ell }\frac{\ell !}{4k_{B}T^{\ell +1}}
\notag \\
& \times \sum_{s=\pm 1}s\int d^{3}k\left( \frac{\partial \varepsilon _{s}}{%
\partial k_{b}}\right) ^{\ell }\left( \varepsilon _{s}-\mu \right) ,
\label{s1}
\end{align}%
where%
\begin{equation}
\varepsilon _{s}=H_{\text{kine}}\left( \mathbf{k}\right) +sJf_{X}\left( 
\mathbf{k}\right) 
\end{equation}%
is the energy for spin $s=\pm 1$. Eq.(\ref{s1}) is Eq.(\ref{Ml}) in the main
text.

 Next, we consider the case $\partial _{x}^{2}T\neq 0$. We expand the
Fermi distribution, $f=f^{\left( 0\right) }+f^{\left( 1\right) }+\cdots $,
and we obtain%
\begin{equation}
\frac{1}{\hbar }\frac{\partial \varepsilon \left( \mathbf{k}\right) }{%
\partial \mathbf{k}}\cdot \frac{\partial f^{\left( \ell -1\right) }}{%
\partial \mathbf{r}}=-\frac{f^{\left( \ell \right) }}{\tau }.
\end{equation}%
It is straightforward to solve the Boltzmann equation recursively and derive
the formula%
\begin{equation}
f^{\left( \ell \right) }=\left( -\frac{\tau }{\hbar }\frac{\partial
f^{\left( 0\right) }}{\partial x}\right) ^{\ell }\left( \frac{\partial
\varepsilon }{\partial k_{b}}\right) ^{\ell },  \label{fl2}
\end{equation}%
where we have assumed that temperature gradient is along the $x$-axis. We
use it to derive Eq.(\ref{paraM}) in the main text.

\end{document}